\documentclass[12pt]{article}
\usepackage[utf8]{inputenc}
\usepackage{varwidth}
\usepackage{cite}
\usepackage{physics}
\usepackage{url}
\usepackage{empheq}
\usepackage{wasysym}
\usepackage{fontawesome}
\usepackage{marvosym}
\usepackage{adforn}
\usepackage{halloweenmath}
\usepackage{staves}
\usepackage{bclogo}
\usepackage{amsthm,amsmath,latexsym,amssymb,amsfonts,amsbsy}
\usepackage{graphicx,lscape,fancyhdr,xcolor,array,stmaryrd,euscript,wrapfig}
\usepackage{hyperref}
\usepackage{tikz}
\usepackage{pgfplots}
\usetikzlibrary{decorations.pathmorphing}
\definecolor{lightblue}{rgb}{0.0, 0.44, 1.0}
\definecolor{purple}{rgb}{0.5,0.0,0.5}
\definecolor{wine}{rgb}{0.59, 0.0, 1.0}
\definecolor{darkmagenta}{rgb}{0.56, 0.0, 0.55}
\definecolor{darkblue}{rgb}{0.0, 0.0, 0.55}
\definecolor{darkred}{rgb}{0.7, 0.0, 0.3}
\hypersetup{
	unicode,	
	colorlinks,
	citecolor=darkblue,
	linkcolor=darkred,
	urlcolor=darkred,
	bookmarksopen=true,
	bookmarksnumbered
	}

 \csname
@addtoreset\endcsname{equation}{section}
\newcommand{\ext}{{\rm d}}
\begin{document}
\begin{titlepage}
\setcounter{page}{1}
\begin{center}
\hfill
\vskip 1pt
{\huge Reconstructing the boundary of AdS\\[5pt] from an infrared defect}
\vskip 30pt
{\sc Cesar Arias}
\vskip 10pt
{
Departamento de Matem\'atica, Pontificia Universidad Cat\'olica de Chile\footnote{Current affiliation. E-mail: \href{mailto:cesar.arias@uc.cl}{cesar.arias@uc.cl}}
\vskip -2pt
and 
\vskip -2pt
Department of Mathematics, University of California, Davis
}
\vskip 30pt
{\bf Abstract}
\end{center}
We argue that the boundary of an asymptotically anti-de Sitter (AdS) space of dimension $d+1$, say~$M^{d+1}$, can be locally reconstructed from a codimension-two defect located in the deep interior of a negatively curved Einstein manifold $X^{d+2}$ of one higher dimension.
This means that there exist two different ways of thinking about the same $d$-submanifold, $\Sigma^d$: either as a defect embedded in the interior of $X^{d+2}$, or as the boundary of $M^{d+1}$ in a certain zero radius limit. 
Based on this idea and other geometric and symmetry arguments, we propose the existence of an infrared field theory on a bulk $\mathbb Z_n$-orbifold defect, located in the deepest point of the interior of AdS$^{d+2}$.
We further conjecture that such a theory gives rise to the holographic theory at the asymptotic boundary of~AdS$^{d+1}$, in the limit where the orbifold parameter~$n\to\infty$.
As an example, we compute a defect central charge when $\Sigma$ is a 2-manifold of fixed positive curvature, and show that its $n\to\infty$ limit reproduces the central charge of  Brown and Henneaux.

\end{titlepage}

{\small\tableofcontents }
\newpage
\section{Introduction}
\subsection{Bulk defects as generalized boundaries}
It has increasingly become a known fact that in order to fully characterize a quantum field theory one should consider not only local operators but also take into account defects of various codimensions. 
A codimension-$k$ defect is a $d$-dimensional submanifold with singular support, embedded in a manifold of dimension $D>d$, where $k=D-d$. Examples include line and surface defects, such as Wilson and 't~Hooft loops and surfaces, and cosmic strings and membranes.

The properties of defects have shown to be relevant in the study of dualities in supersymmetric gauge theories~\cite{Alday:2009aq,  Alday:2009fs, Drukker:2009id, Drukker:2010jp}, boundary conformal field theories~\cite{Cardy:1984bb,McAvity:1993ue,McAvity:1995zd,Takayanagi:2011zk,Herzog:2015ioa,Fursaev:2015wpa,Solodukhin:2015eca, Billo:2016cpy, Herzog:2017xha}, and in the study of generalized symmetries and charges in field theory~\cite{Gaiotto:2014kfa, Albertini:2020mdx, Morrison:2020ool} and higher spin gravity~\cite{Colombo:2012jx, Bonezzi:2017vha}. 
The case of orbifold defects has been of importance in the computation of holographic R\'enyi and entanglement entropies~\cite{Ryu:2006bv, Dong:2016fnf,Lewkowycz:2013nqa}, and in the analysis of the Page curve of evaporating black holes~\cite{Penington:2019npb, Almheiri:2019psf,Almheiri:2019hni,Rozali:2019day}.

The aim of this article is to argue that, on general grounds, a bulk defect\footnote{In this work, we are interested in defects that are located in the interior of a manifold.  A defect with support on a boundary subregion is sometimes referred to as a corner.} and a boundary are two different phases of the \emph{same} object; a $d$-submanifold, say $\Sigma^d$, can be understood as a defect or as a boundary depending on the different limits of the theory one is looking at. 
Moreover, as we will elaborate on for the case of an asymptotically AdS space $M^{d+1}$, the boundary submanifold can be \emph{reconstructed} from a bulk defect embedded in a manifold of one higher dimension, that hereafter we denote by $X^{d+2}$.  The general scheme is illustrated in following diagram\footnote{
We decorate the manifold $\Sigma_\star^d$ with a ``star" to specify that is being treated as a defect; we write $\Sigma_\star^d\hookrightarrow X^{d+2}$ to indicate that the defect $\Sigma_\star^d$ is embedded in $X^{d+2}$. When the manifold instead behaves as a boundary, we simply write $\Sigma^d=\partial M^{d+1}$ with no extra bells or whistles.  The rest of the notation used through the paper is collected in appendix \ref{appA}.
}:
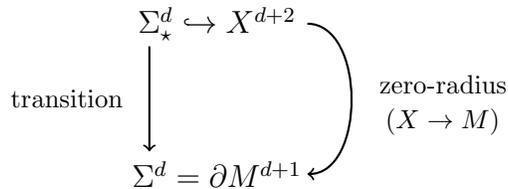
\bigskip
\begin{center}
\begin{tikzpicture}
\node at (0,2) {$\Sigma_\star^d\hookrightarrow X^{d+2}$};
\node at (-2,1) {\footnotesize transition};
\node at (0,0) {$\Sigma^d=\partial M^{d+1}$};
\draw[thick, ->](-0.9,1.7)--(-0.9,0.3);
\draw[thick, ->] (1.2, 2)to[out=0,in=0](1.2, 0);
\node at (3,1.2) {\footnotesize zero-radius};
\node at (3,0.7) {\footnotesize($X\to M$)};
\end{tikzpicture}
\end{center}

In this picture, all the geometric properties of the asymptotic AdS boundary may be thought of as being inherited from a higher (co)dimensional bulk defect, as a result of some type of transition $\Sigma^d_\star\to\Sigma^d$ whereby the boundary $\Sigma^d=\partial M^{d+1}$ is truly a reincarnation of a defect $\Sigma^d_\star$ embedded in $X^{d+2}$.
The existence of this transition yields inevitably to hypothesize that bulk defects should be able to holographically encapsulate~\cite{tHooft:1993dmi, Susskind:1994vu} (just as a boundary does) degrees of freedom that can independently be described by means of some field theory; we conjecture that such a theory gives rise to the holographic theory at the boundary of AdS~\cite{Maldacena:1997re, Gubser:1998bc, Witten:1998qj}, in a certain zero-radius limit. 


\subsection{Summary and plan of the paper}
Having in mind the diagram displayed further above, we begin in~\S\ref{sec2} by constructing the manifold~$X^{d+2}$. For simplicity, we take $X^{d+2}=D^2\times \Sigma^d$, where  $D^2$ is a disk (with boundary a circle) and $\Sigma^d$ is a $d$-dimensional manifold with no boundary (that we can think of as having sphere topolgy). Importantly,  one can easily create a defect on $X^{d+2}$ by acting with $\mathbb Z_n$ on the disk; this originates a defect in codimension two, that we denote by $\Sigma_\star$, which corresponds to the set of fixed points of the $\mathbb Z_n$ action on $X^{d+2}$ and it is thus located at the center of the disk. This is of course the deepest point of the interior of $X^{d+2}$.

We next ask ourselves whether if physically relevant spacetimes of this type can actually exist; requiring $X^{d+2}$ to be a negatively curved Einstein manifold, we show the existence of a family of such backgrounds---all of them supporting a defect $\Sigma_\star$ in the deepest point of their interior---, of which pure AdS spacetime is an example. 
Motivated by the ideas of the holographic renormalization group flow~\cite{Akhmedov:1998vf, Alvarez:1998wr, Girardello:1998pd, Distler:1998gb, Balasubramanian:1999jd, Porrati:1999ew, Girardello:1999bd, Skenderis:1999mm, deBoer:1999tgo}, in which one identifies the AdS radius as the energy scale in the flow of the dual field theory, we refer to $\Sigma_\star$ as an infrared defect. 

In \S\ref{sec3} we study the local geometry close to an infrared defect by zooming into the region at center of the disk. About this region, the quotient $D^2/\mathbb Z_n$ is locally a cone, and the manifold $X^{d+2}$ is approximately the direct product of that cone (with the defect $\Sigma^d_\star$ at the tip of it) with $\Sigma^d$; see figure~\hyperlink{Fig:1}1.
Importantly, the radius of the cone scales as $1/n$, where $n>1$ is the $\mathbb Z_n$-orbifold parameter. Thus, the limit\footnote{Here and in what follows, we implicitly assume the analytic continuation $n\in\mathbb R_+$.} $n\to\infty$ is equivalent to the zero-radius limit of the cone. In this limit, the cone shrinks to a small interval, say $[0, \varepsilon)$ (where $\epsilon>0$ defines the range of validity of the local approximation), and thus the full space $X^{d+2}$ collapses to $M^{d+1}= [0, \varepsilon)\times\Sigma^d$. During the process, the defect submanifold~$\Sigma^d_\star$---originally embbeded in $X^{d+2}$---becomes the boundary of $M^{d+1}$, as illustrated in the diagram of the previous page.  We denote this transition\footnote{In two bulk dimensions, related ideas have been explored in string theory and condensed matter physics when studying the behavior of boundary degrees of freedom under renormalization group flow~\cite{Dorey:2000eh, Keller:2007nd, Hori:2004bx, Fredenhagen:2006dn, Brunner:2007ur, Gaiotto:2012np}.} as $\Sigma^d_\star\to\Sigma^d$.

We continue by observing that the product $[0, \varepsilon)\times\Sigma^d=M^{d+1}$, where we recall that $\Sigma^d$ has no boundary, has the same form as the collar neighborhood one considers when studying the geometry close to the boundary of an asymptotically AdS space. Therefore, in \S\ref{sec32} we ask for the conditions under which the manifold $M^{d+1}=X^{d+2}\vert_{n\to\infty}$ (understood as a limit of $X$) can be regarded as an asymptotically AdS space.  
These conditions follow from requiring that Einstein's equation for the metric close to the boundary $\Sigma^d=\partial M^{d+1}$---which can in general be solved asymptotically by means of the Fefferman--Graham expansion~\cite{Fefferman:2007rka}---should arise from the large~$n$ limit of Einstein's equations for the metric on $X^{d+2}$, about the region close to $\Sigma^d_\star$. We will refer to the procedure of imposing such conditions as \emph{boundary reconstruction}. 
 
Next, in \S\ref{sec33}, we show that, just as in the case of the boundary of AdS, the Einstein condition at finite $n>1$, on the metric close to the defect, can also be formally solved order by order in powers of the distance to the defect, in a metric expansion that resembles the Fefferman--Graham solution. We construct this expansion up to second order.

In \S\ref{sec4} we turn to the holographic implications of the $\Sigma^d_\star\to\Sigma^d$ transition; because of the existence of a dual theory at the boundary $\Sigma^d$ of an asymptotically AdS space~\cite{Maldacena:1997re, Gubser:1998bc, Witten:1998qj}, it is plausible to think that (at least in some cases) such a theory exists already on an infrared defect~$\Sigma^d_\star$, and becomes a boundary theory only in the zero-radius limit $n\to\infty$.
Consequently, in \S\ref{sec41}, we argue that the parent bulk defect~$\Sigma^d_\star$ exhibits generalized versions of all the relevant features that we find at the asymptotic boundary of AdS, and that are indicative of the existence of a boundary holographic theory, namely:
\begin{itemize}
\item[$\diamond$] At the location of the defect, the spacetime symmetries are enhanced to those of the full conformal group.
\item[$\diamond$] The singular nature of the defect permits the insertion---via a $\delta$-function in codimension-two---of a local stress-energy tensor, which in principle suffices to define a conformal field theory.
\item[$\diamond$] In a suitable gauge, the defect submanifold turns out to be naturally equipped with a conformal equivalence class of metrics (also known as a conformal structure). Furthermore, in that gauge and as spelled out in \S\ref{sec33}, there exists a formal asymptotic solution to Einstein's equations for the metric about the location of the defect, analogue to the Feffereman--Graham expansion, whose expansion coefficients encode relevant holographic quantities. 
\end{itemize}

The previous elements lead us to propose the existence of a conformal field theory (CFT) on a $\mathbb Z_n$-orbifold defect $\Sigma_\star^d$, located in the deepest point of the interior of $X^{d+2}$. 
We further argue that, by virtue of the $\Sigma^d_\star\to\Sigma^d$ transition, this theory gives rise to the holographic theory at the boundary $\Sigma^d$ of an asymptotically AdS space $M^{d+1}$, upon taking the $n\to\infty$ limit.

In \S\ref{sec42} we give a simple example.
We compute a central charge for the theory on the infrared defect in the case where the defect is a 2-manifold embedded in four bulk dimensions.
We show that when $\Sigma^2_\star$ has scalar Ricci curvature equal to $4/R_0^2$, where $R_0$ is the radius of the disk, the $n\to\infty$ limit of the charge reproduces the Brown--Henneaux~\cite{Brown:1986nw} central charge of the theory at the asymptotic boundary of AdS$^3$.

We conclude with a brief discussion in \S\ref{discussion}. 
We collect our conventions, notation, and some details of our calculations in appendices \ref{appA} and \ref{appB}.

The main ideas presented here have been shaped by previous work~\cite{ Arias:2019zug, Arias:2019pzy}, in which the properties of defects in codimension-two were exploited to study the notion of entanglement in de Sitter space (see also~\cite{Arias:2012dj} for earlier work dealing with defects in de Sitter space). In particular, the idea that a boundary submanifold can be understood as a limit of a bulk defect was first discussed in~\cite{Arias:2019zug}. In a more formal context, some of the holographic features of embedded submanifolds in arbitrary codimension were studied in~\cite{Arias:2019mzc}.
Different routes to generalize holography to higher codimensions, in which the dual CFT has support on a boundary corner, have been proposed by other authors in~\cite{Akal:2020wfl, Miao:2020oey, Geng:2020fxl, Miao:2021ual}.

\section{Global Einstein geometries with defects} 
\label{sec2}
In this section, we study a certain type of Einstein geometries that admit a codimension-two defect in the deepest point of their interior. We demonstrate the existence of an entire family of such geometries and, as an example, we explicitly show that pure AdS spacetime belongs to this family. 
The goal of this section is to motivate a further local, asymptotic analysis about the location of one of these defects, in the same fashion one performs a local study about the boundary of an asymptotically AdS spacetime. 

\subsection{Geometries with a deep-in-the-bulk defect}
\label{sec21}
To begin with, we consider a manifold $X$ of dimension $d+2$ given by the direct product
\begin{equation}\label{X}
X^{d+2} = D^2/\mathbb Z_n \times \Sigma^d~,
\end{equation}
where $D^2$ is a two-dimensional disk and $\Sigma^d$ is a $d$-dimensional manifold without boundary. It follows that\footnote{Whenever is clear from context, we will drop the dimension as a superscript and simply write $X$ and $\Sigma$ instead of $X^{d+2}$ and $\Sigma^d$.}  
\begin{equation}\label{partialX}
\partial X = S^1_{r_n} \times \Sigma~,
\end{equation}
where $S^1_{r_n}$ denotes a circle of radius $r_n$; due to the $\mathbb Z_n$ action on $D^2$, this radius scales as $r_n\sim 1/n$.
We next endow $X$ with a singular metric of the form
\begin{equation}\label{gX}
g_X =  \frac{g_{D^2/\mathbb Z_n} + h}{u^2}~.
\end{equation}
Here, $g_{D^2/\mathbb Z_n}$ is a two-dimensional Euclidean metric on the conically singular orbifold $D^2/\mathbb Z_{n}$ (whose smooth limit is $n=1$), $h$ is a Lorentzian metric on $\Sigma^d$, and $u$ is a defining function whose zero locus determines the conformal infinity of the metric~\eqref{gX}, that is
\begin{equation}\label{confinf}
{\rm Conf}_\infty (g_X) := \{ p\in X\vert~u(p)=0 \}~.
\end{equation}

Choosing the coordinates on $D^2$ to be $(\theta, \phi)$, with $0\leq\theta\leq\pi/2$ and $0\leq\phi<2\pi$, the coordinates on $\Sigma$ to be $x^i$, with $i=0,...,d-1$, and recalling that $\mathbb Z_n$ acts on $D^2$ by the azimuthal identification $\phi\sim\phi+2\pi n^{-1}$, the metric~\eqref{gX} reads
\begin{equation}\label{gXcoords}
g_X = \frac{R_0^2(\ext \theta^2 + n^{-2}\sin^2\theta\,\ext\phi^2)+h_{ij}(\theta, x)\ext x^i\ext x^j}{u^2(\theta)}~. 
\end{equation}
In the above, $R_0$ denotes the radius of the disk, and we have taken the metric $h$ to depend on $x\in\Sigma$ and on the polar coordinate $\theta\in D^2$, while the defining function $u$ only depends on the latter.

We are interested in the case in which the pair $(X, g_X)$ is an Einstein manifold for a negative cosmological constant. When that is the case, Einstein equations for the components $(g_X)_{\theta \theta}$, $(g_X)_{\theta i}$, $(g_X)_{\phi\phi}$ and $(g_X)_{ij}$ are respectively given by\footnote{We have retained the overall factor of $1/n^2$ in front of~\eqref{Eeff} for reasons that will become clear later.}
\begin{align}
\label{Eett}
0&=-\frac{1}{2}{\rm Tr}(h^{-1}h'')+\frac{1}{4}{\rm Tr}(h^{-1}h'h^{-1}h')+\frac{1}{2}\frac{u'}{u}{\rm Tr}(h^{-1}h')\cr
&\hspace{6.5cm}+1+\cot\theta\,\frac{u'}{u}+(d+1)\bigg[\frac{u''}{u}-\bigg(\frac{u'}{u}\bigg)^2+\frac{R_0^2}{L^2 u^2}\bigg]~,\\
\label{Eeti}
0&=h^{jk}(\nabla_j h'_{ik} -\nabla_i h'_{jk})~,\\
\label{Eeff}
0&=\frac{\sin^2\theta}{n^2}\bigg[1+\frac{u''}{u}+\bigg(\frac{u'}{u}-\cot\theta\bigg)\bigg(\frac{1}{2}{\rm Tr}(h^{-1}h')-(d+1)\frac{u'}{u}\bigg)+\frac{(d+1)R_0^2}{L^2u^2}\bigg]~, \\ 
\label{Eeij}
0&=R_{ij}(h)-\frac{1}{2R_0^2}h''_{ij}  +\frac{1}{2R_0^2} (h'h^{-1}h')_{ij}- \frac{1}{2R_0^2}\bigg(\cot\theta-d\,\frac{u'}{u}+\frac{1}{2}{\rm Tr}(h^{-1}h')\bigg) h'_{ij}\cr
&\hspace{4cm}+\frac{1}{R_0^2}\bigg[\frac{u''}{u}+\cot\theta\,\frac{u'}{u} +\frac{d+1}{u^2}\bigg(\frac{R_0^2}{L^2}-(u')^2\bigg)+\frac{1}{2}\frac{u'}{u}{\rm Tr}(h^{-1}h')\bigg] h_{ij}~.
\end{align}
The calculation of the above equations makes use of the conventions specified in appendix~\ref{appA} and the components of the Ricci tensor for $g_X$ given in equation~\eqref{RicX}. In order to lighten the notation, we have suppressed whenever is possible the indexes on the metric $h$, writing ${\rm Tr}(h^{-1}h'')=h^{ij}h''_{ij}$, ${\rm Tr}(h^{-1}h'h^{-1}h')=h^{ij}h'_{ik}h^{kl} h'_{jl}$, and $(h'h^{-1}h')_{ij}=h'_{ik}h^{kl}h'_{jl}$, where the primes indicate derivatives with respect to~$\theta$. Also, $L$ denotes the AdS$^{d+2}$ radius, and $R_{ij}(h)$ denotes the components of the Ricci tensor built from $h$. 

Global solutions to \eqref{Eett}-\eqref{Eeij} are difficult to find, of course, and it is not our purpose here. However, we observe that a family of exact Einstein geometries can be obtained by taking $h$ to be independent of~$\theta$ (so that $h'=h''=0$), and by setting the radius of the disk to be equal to the AdS$^{d+2}$ radius, that is
\begin{equation}
h=h(x)~~~{\rm and}~~~R_0= L~. 
\end{equation}
Consequently, the defining function
\begin{equation}\label{u}
u=\cos\theta
\end{equation}
solves equations~\eqref{Eett}-\eqref{Eeij} in the special case in which $h$ is itself \emph{any} Einstein metric of negative scalar of curvature, namely
\begin{equation}\label{Econst}
R_{ij}(h) + \frac{d-1}{L^2} \, h_{ij} =0~.
\end{equation}
%

We thus have the family of Riemannian geometries 
\begin{equation}\label{Fh}
\mathcal F_{h}:=(X, g_X(h))~,
\end{equation} 
where $X$ and $g_X$ are defined as in~\eqref{X} and~\eqref{gXcoords}, respectively, and $h$ satisfies the Einstein condition~\eqref{Econst}. From~\eqref{confinf} and~\eqref{u}, it follows that in all these geometries the conformal infinity of $g_X$ is located at $\theta=\pi/2$ and thus coincides with the boundary of $X$:
\begin{equation}
{\rm Conf}_{\infty} (g_X) = \partial X~.
\end{equation}

A key feature of the family $\mathcal F_h$ is that every geometry member of it contains two distinguished submanifolds, namely a codimension-one boundary and a codimension-two bulk defect. Indeed, recalling that $X^{d+2}\cong D^2/\mathbb Z_n \times \Sigma^d$, one can see that the non-trivial $\mathbb Z_{n>1}$ action has as a set of fixed points the center of the disk (see figure~\hyperlink{Fig:1}1), which is the deepest point of the interior of~$X$ (i.e. the furthest point from the boundary). 
Metric-wise and locally about this point, in coordinates given by $\theta=0$, we have that $g_X\approx R^2_0(\ext \theta^2 + \theta^2\ext\phi^2/n^2)+\cdots$, which is the singular geometry of a cone of deficit angle $2\pi(1-1/n)$.
\vspace{1cm}
\begin{center}
\begin{tikzpicture}[scale=0.8]
\node at (-3,2.7) {Smooth geometry ($n=1$)};
\node at (-3,-0.8) {$\partial X={\rm Conf}_\infty (g_X)$};
\draw [thick, darkred](-3,0) ellipse (2cm and 0.3cm);
\draw[thick] (-1,0) 
arc[start angle=0,end angle=-180, 
x radius=2, y radius=-2];
\coordinate (N) at (4,2);
\coordinate (E1) at (2.5,0);
\coordinate (E2) at (5.5,0);
\node at (4,2.7) {$\Sigma_\star$ defect ($n>1$)};
\draw [thick, darkred](4,0) ellipse (1.5cm and 0.3cm);
\draw[thick] (N)to[out=-20,in=100](E2);
\draw[thick] (N)to[out=-160,in=80](E1);
\node at (4.2,-0.8) {$\partial  X$};
\node  at (N) {\color{darkred}\textbullet};
\draw [thick,darkblue,->,>=latex] (4.1,0) -- (5.5,0);
\node at (6.9,0) {\color{darkblue}$r_n=R_0/n$};
\node[text width=15cm, text justified] at (0.7,-5.2){
{\hypertarget{Fig:1} \bf Fig.\,1}: 
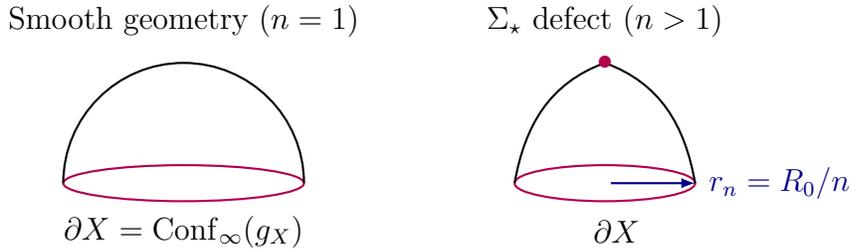
Depiction of the $D^2/\mathbb Z_n$ factor of the product manifold $X^{d+2}=D^2/\mathbb Z_n\times\Sigma^d$. For any $h$ satisfying the Einstein condition~\eqref{Econst}, the corresponding member of the family $\mathcal F_h=(X, g_X(h))$ has conformal infinity at the boundary of the disk, which coincides with the boundary of $X$.  On the left, we illustrate the smooth, $n=1$ geometry, in which case the radius of the disk is $R_0$.
The value $n>1$, on the right, induces a conical defect $\Sigma_\star$ (the set of fixed points of the $\mathbb Z_n$-action) located at the center of the disk, which is the deepest point of bulk. In this case, the boundary circle has radius $r_n=R_0/n$. };
\end{tikzpicture}
\end{center}

In what follows, we will denote the codimension-two set of fixed points as
\begin{equation}\label{SX0}
\Sigma_\star := X\big\vert_{\theta=0}~,
\end{equation}
and we will refer to it as \emph{defect}. By construction, $\Sigma_\star$ has the same topology as $\Sigma$ (and hence has no boundary), and it is endowed with the induced metric $h_{(0)}=g_X\vert_{\theta=0}$ that in turn satisfy~\eqref{Econst}.

\subsection{Example: pure AdS}
\label{sec22}
There exists a distinguished solution $h$ to the Einstein condition~\eqref{Econst} by means of which the Riemannian manifold $(X^{d+2}, g_X)$ turns into pure AdS$^{d+2}$ spacetime. 
To this end, we take $\Sigma^d$ to be two copies of AdS$^d$ glued along their boundaries, that is 
\begin{equation}\label{AdSpm}
\Sigma^d = {\rm AdS}^d_{\pm}:= {\rm AdS}^d_{+} \cup {\rm AdS}^d_{-}~,
\end{equation}
where we have denoted by ${\rm AdS}^d_{+}$ and ${\rm AdS}^d_{-}$ to each of these copies. Note that since the gluing is along the boundary, $\Sigma$ has sphere topology and thus no boundary.
We next equip AdS$^d_\pm$ with the line element
\begin{equation}\label{hpm}
h=h_{{\rm AdS}_{\pm}^d} = \frac{L^2}{\cos^2 z}\Big[-\ext t^2+\ext z^2+\sin^2z\,\ext\Omega^2_{d-2}\Big] ~.
\end{equation} 
These coordinates are sometimes referred to as the conformal compactification of AdS; in our case, the radial coordinate $0\leq z\leq \pi$, with $0\leq z\leq\pi/2$ for one AdS copy and $\pi/2\leq z\leq \pi$ for the second one. The two copies are glued along the boundary\footnote{In the special case of AdS$^2$ the gluing is made along the two disconnected boundaries located at $z=0,\pi$; the resulting extended $z$-coordinate runs then over an entire circle $0\leq z<2\pi$.} located at $z=\pi/2$. As usual, the time coordinate $-\infty<t<\infty$, and $\ext\Omega^2_{d-2}$ denotes the induced metric on a sphere of dimension~$d-2$.

With the choices~\eqref{AdSpm} and~\eqref{hpm}, the full $(d+2)$-dimensional geometry~\eqref{gXcoords} becomes
\begin{equation}\label{g1}
g_X=\frac{L^2(\ext \theta^2 + \sin^2\theta\,\ext\phi^2) + h_{{\rm AdS}_\pm^d}}{\cos^2\theta}~.
\end{equation} 
A direct calculation shows that~\eqref{g1} is indeed the induced metric on the 
AdS$^{d+2}\hookrightarrow \mathbb R^{2,d+1}$ hyperboloid
\begin{equation}\label{hyper}
-(Z^{0})^2 - (Z^{0'})^2+ \sum_{a=1}^{d+1} (Z^a)^2 = -L^2~,
\end{equation}
where the $Z$'s are coordinates on flat embedding space $\mathbb R^{2, d+1}$. To see this, it suffices to parametrize
\begin{align}\label{Xs}
&Z^0= \frac{L\cos(t/L)}{\cos\theta\cos z}~, \qquad
Z^{0'}= \frac{L\sin(t/L)}{\cos\theta\cos z}~,\cr
Z^i=\frac{L\tan z }{\cos\theta}\,&y^i~, \qquad
Z^d=L \tan\theta\cos\phi~,\qquad
Z^{d+1}=L \tan\theta\sin\phi~,
\end{align}
where we recall that  $0\leq\theta\leq\pi/2$ and $0\leq\phi\leq2\pi$.  
The pullback of the flat embedding space metric $\eta={\rm diag} (-1,-1,1,...,1)$ onto the hypersurface~\eqref{hyper} then gives~\eqref{g1}.  

\section{Boundary reconstruction and local defect geometry}
\label{sec3}
We now abandon the global approach of~\S\ref{sec2} and focus on the asymptotic geometry about the region close to the defect.  Our first goal here is to examine, locally around $\Sigma_\star:= X\big\vert_{\theta=0}$, the $n\to\infty$ limit of equations~\eqref{Eett}-\eqref{Eeij}. 
This limit---which corresponds to the zero-radius limit of an azimuthal circle transverse to $\Sigma_\star$---defines a transition whereby the defect submanifold, originally embedded in $X^{d+2}$, reincarnates as the boundary of the resulting space of one lower dimension, that we denote by $M^{d+1}$; the situation is depicted down below.
\vspace{1cm}
\begin{center}
\begin{tikzpicture}
\draw[thick] (0,0) -- (2,1);
\draw[thick] (0,0) -- (2,-1);
\node [darkred] at (0,0) {\textbullet};
\node at (-1,0.2) {$\Sigma_\star\hookrightarrow X$};
\node at (0,-1.5) {(a) Finite $n>1$};
{\color{darkblue}
\draw[thick, dashed] (1.7,0) 
arc[start angle=0,end angle=180, 
x radius=0.2, y radius=0.75];
\draw[thick, dashed] (1.7,0) 
arc[start angle=0,end angle=-180, 
x radius=0.2, y radius=0.75];
}
\draw[thick] (5,0) -- (7.5,0);
\node [darkred] at (5,0) {\textbullet};
\node at (5,0.5) {$\Sigma=\partial M$};
\node at (6,-1.5) {(b) The limit $n\to\infty$};
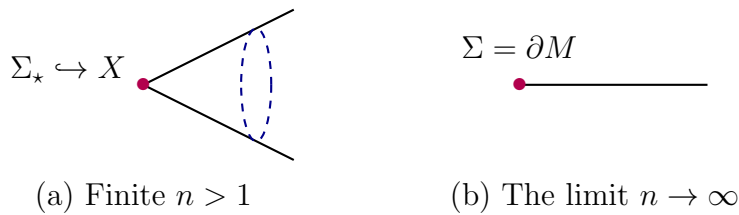
\node[text width=14cm, text justified] at (2.7,-4){{\hypertarget{Fig:2}\bf Fig.~2}: Local picture of the quotient $D^2/\mathbb Z_n$ about the center of the disk: (a) for finite $n>1$, the set of fixed points $\Sigma_\star$ is a codimension two defect embedded in $X$; (b) when $n\to\infty$, the transverse circle shrinks to a point and $D^2/\mathbb Z_n$ collapses to the interval $I$. During the process, $\Sigma_\star$ transitions from being a defect embedded in $X^{d+2}$, to be the boundary of $M^{d+1}\cong I\times\Sigma^d$.
};
\end{tikzpicture}
\end{center}
Consequently, in \S\ref{sec32}, we establish the conditions under which $M^{d+1}$ can be generically considered an asymptotically AdS spacetime, with its ordinary asymptotic boundary $\Sigma=\partial M$ being thought of as the large $n$ phase of the defect, finite $n>1$ submanifold $\Sigma_\star$. 

The second goal of the section is to show that the local defect equations, given by the $\theta\ll1$ approximation of equations~\eqref{Eett}-\eqref{Eeij}, can be formally solved order by order in the ``radial" coordinate, in a metric expansion that resembles the Fefferman--Graham boundary expansion; we construct such an expansion up to second order in~\S\ref{sec33}.
Because it is needed for our purposes, we begin in \S\ref{sec31} by reviewing the relevant properties of the asymptotic geometry of the boundary of AdS.

\subsection{The local geometry of the AdS boundary}
\label{sec31}
In~\cite{FG}, Fefferman and Graham established a link between a pseudo\footnote{The prefix \emph{pseudo} here simply means that the (everywhere non-degenerate) metric $g_Y$ needs not to be positive definite, so that is taken to be an \emph{indefinite} bilinear form.}-Riemannian \emph{ambient} manifold $(Y^{d+2}, g_Y)$ of dimension $d+2$, and a conformal manifold $\Sigma^d$ of dimension $d$, by means of which \emph{local conformal invariants} on $\Sigma$ can be constructed from \emph{Riemannian} invariants on $Y$. 
The construction of these invariants is carried out by formally solving a Ricci-flat condition for the ambient space metric; physics-wise, this Ricci-flat condition for $g_Y$ happens to be equivalent to Einstein's equations on a negatively curved manifold $M^{d+1}$ with boundary $\Sigma^d$. Consequently, as suggested in~\cite{Witten:1998qj}, the Fefferman-Graham construction naturally encapsulates some of the geometric textures appearing in Maldacena's AdS/CFT correspondence~\cite{Maldacena:1997re}. In particular, its usage has been relevant to the calculation of the holographic Weyl (boundary) anomaly~\cite{Henningson:1998gx}, as well as other types of submanifold anomalies~\cite{Graham:1999pm}.    
In what follows, we briefly review the aspects of the Fefferman-Graham expansion that are relevant for our purposes. Further details can be found in the monograph~\cite{Fefferman:2007rka}. 

\paragraph{Conformally compact Einstein metrics.}
Let $\overline M= M\cup\,\partial M$ be a compact manifold of dimension $d+1$ with interior $M$ and boundary $\partial M=\Sigma$.
A Riemannian metric $g$ on $M$ is said to be conformally compact if there exists a smooth defining function $r\in\mathcal C^{\infty}(\overline M)$, in which case
\begin{equation}
r \vert_{M} >0~,\quad
r\vert_{\Sigma} =0~,\quad 
\ext r\vert_{\Sigma}\neq 0~,
\end{equation}
such that the metric   
\begin{equation}
\overline g= r^2 g
\end{equation} 
extends continuously to $\overline M$. The structure $(\overline M, \overline g)$ is referred to as a compactification of $(M, g)$~\cite{Penrose:1986ca}. Because the choice of defining function is not unique, the restriction $h_{(0)}$ of $\overline g$ to $\partial M$ rescales upon different choices of $r$; this freedom invariantly defines a conformal class of metric $[h_{(0)}]$ on $\partial M$. The pair $(\Sigma, [h_{(0)}])$ is the conformal infinity of the metric $g$.

A metric $g$ which in addition satisfies the Einstein condition $\ell^2 R_{ij}(g)+dg=0$, where $\ell$ is the radius of curvature of the manifold $M$, is termed a conformally compact Einstein metric. Importantly, every conformally compact Einstein metric is asymptotically hyperbolic\footnote{In order to be consistent with the original literature, in this subsection we are considering spaces of Euclidean signature; statements regarding asymptotically hyperbolic spaces translate with no subtleties to asymptotically AdS spaces in Lorentzian signature.}, meaning that its sectional curvature approach to $-1/\ell^2$ at $\Sigma$. Sometimes in the math literature thess type of metrics are dubbed Poincar\'e--Einstein metrics.
\paragraph{Graham--Lee normal form and Fefferman--Graham expansion.}  If $g$ is an asymptotically hyperbolic metric on $M$, then a choice of a representative $h_{(0)}$ in the conformal class $[h_{(0)}]$ on $\Sigma$ uniquely determines a defining function $r$ such that, in a collar neighborhood $\Sigma\times [0,\varepsilon)$, the singular metric $g$ takes the Graham--Lee normal form~\cite{Graham:1991jqw}
\begin{equation}\label{normal}
g=\frac{\ell^2(\ext r^2 + h_r)}{r^2}~,
\end{equation}
where $h_r$ is a one-parameter family of metrics on $\Sigma$, with $h_0=h_{(0)}\in [h_{(0)}]$.

The Einstein condition $\ell^2 R_{ij}(g)+dg=0$ can be asymptotically solved for a metric of the form~\eqref{normal}. The solution is a formal expansion $h_r=\sum_{k\geq0} h_{(k)}r^k$, where the expansion coefficients $h_{(k)}$ are determined inductively from the Einstein condition itself. In components, this condition reads
\begin{align}
\label{rr}
r{\rm Tr}(h^{-1}h'')-\frac{r}{2}{\rm Tr}(h^{-1}h'h^{-1}h') - {\rm Tr}(h^{-1}h')&=0~,\\
\label{ri}
\nabla_i {\rm Tr}(h^{-1}h')-\nabla^j h'_{ij} &=0~,\\
\label{FGeq}
r h''_{ij} + (1-d)h'_{ij}-{\rm Tr}(h^{-1}h')h_{ij}-r\bigg[(h'h^{-1}h')_{ij} -\frac{1}{2}{\rm Tr}(h^{-1}h')h'_{ij}+2  \ell^2 R_{ij}(h)\bigg]&=0~.
\end{align} 
For simplicity, in the above display we have written the tensor $h_r$ simply as $h$, whose components are denoted by $h_{ij}$; we have also denoted ${\rm Tr}(h^{-1}h'')=h^{ij}h''_{ij}$, ${\rm Tr}(h^{-1}h'h^{-1}h')=h^{ij}h'_{ik}h^{kl} h'_{jl}$, and $(h'h^{-1}h')_{ij}=h'_{ik}h^{kl}h'_{jl}$, where the primes indicate derivatives with respect to~$r$.

Successive derivatives of~\eqref{FGeq} evaluated at $r=0$ give 
\begin{equation}\label{DFG}
\Big[(k-d)\partial_r^k h_{ij} - {\rm Tr} (h^{-1} \partial^k_r h)h_{ij}\Big]\Big|_{r=0} = \text{LOTs}\big|_{r=0}~,
\end{equation}
where ``LOTs" refers to lower order terms in derivatives of the metric $h$. From equation~\eqref{DFG} with $h_{(0)}$ as an initial condition, the higher order coefficients can be iteratively determined as follows:
\begin{itemize} 
\item[$\diamond$] For $k<d$, all the coefficients $h_{(k)}$ can be computed in terms of $h_{(0)}$ from $\partial^k_r h$ evaluated at~$r=0$.  The case $k=1$ implies immediately that the first order expansion coefficient $h_{(1)}$ vanishes. Consequently, since equation~\eqref{FGeq} is invariant under $r\to-r$, it follows that only even powers of the expansion have non-vanishing coefficients and thus $h_{(k)}\sim\partial^k_rh\vert_{r=0}=0$ for $k$ odd. 
\end{itemize}
When $k=d$ the coefficient of the trace-free part of $h_{(d)}\sim\partial^d_rh\vert_{r=0}$ vanishes and this can be freely chosen; this is the second piece of initial data---in addition to $h_{(0)}$---needed to solve the second order Einstein condition.  Furthermore 
\begin{itemize}
\item[$\diamond$] If $k=d$ is odd, the LOTs in~\eqref{DFG} vanish at $r=0$ and thus the trace ${\rm Tr} (h_{(0)}^{-1} h_{(d)})=0$. 
\item[$\diamond$] If $k=d$ is even, the trace-free part of the LOTs in~\eqref{DFG} do not vanish at $r=0$ giving rise to what is known as the \emph{obstruction tensor}.  In order to circumvent this obstruction, one must include in the expansion a logarithmic term $r^d\log r$ with a trace-free coefficient~$a_{(d)}$. 
\end{itemize}
Tying the above arguments together one concludes that
\begin{equation}\label{FGexp}
h_r=\begin{cases}
h_{(0)} + h_{(2)}r^2+\text{even powers}+h_{(d-1)}r^{d-1} + h_{(d)}r^d\cdots , & \text{if $d$ is odd}.\\
h_{(0)} +h_{(2)}r^2+\text{even powers}+ a_{(d)} r^d \log r+h_{(d)}r^d+\cdots, & \text{if $d$ is even}.
\end{cases}
\end{equation}
The distinguished coefficients $a_{(d)}$ and $h_{(d)}$ can be characterized as the metric variation of the conformal anomaly, and the expectation value of the boundary stress-energy tensor, respectively.

As for the other two components~\eqref{rr} and~\eqref{ri} of the Einstein condition, it can be shown that they give no extra information at order $k=d$ and lower. This is because, from the ambient space perspective, some of the components of the Ricci-flatness equation for the ambient space metric~$g_Y$ are identically satisfied due to the contracted Bianchi identities~\cite{FG}.

\paragraph{Second–order coefficient.}  For the sake of completeness, let us compute $h_{(2)}$. Evaluating equation~\eqref{FGeq} at $r=0$ implies that $h_{(1)}=0$; using this fact, the first derivative of that equation gives
\begin{equation}\label{h2app}
\Bigg[(2-d)h''_{ij} - {\rm Tr}(h^{-1}h'')h_{ij} -2 \ell^2 R_{ij}(h)\Bigg]_{r=0}=0~.
\end{equation}
Since $h\vert_{r=0}=h_{(0)}$ and $h''\vert_{r=0}= 2h_{(2)}$ (because $a_{(2)}$ is traceless), it follows that for $d=2$ we can only determine the trace of the second-order coefficient
\begin{equation}\label{2ndorder}
{\rm Tr}(h_{(0)}^{-1}h_{(2)})=-\frac{\ell^2}{2} R_{\Sigma}~,
\end{equation}
where $R_{\Sigma}= R(h)\vert_{r=0}$ is the Ricci scalar of the boundary (built from the induced metric $h_{(0)}$). 
It is not difficult to check that when $d>2$, taking the trace of~\eqref{h2app} and plugging it back, one obtains
\begin{equation}
h^{(2)}_{ij}= \frac{\ell^2}{2-d} \bigg[R^{\Sigma}_{ij} - \frac{R_{\Sigma}}{2(d-1)}h^{(0)}_{ij}\bigg] = -\ell^2 P^{\Sigma}_{ij}~, 
\end{equation}
where $P^{\Sigma}_{ij}$ is the Schouten tensor of the boundary.

\subsection{Boundary reconstruction and the Fefferman--Graham--Lee limit}
\label{sec32}
As anticipated, the limit $n\to\infty$ defines the transition $\Sigma_\star\to\Sigma$ in which the defect submanifold $\Sigma_\star\hookrightarrow X^{d+2}$ becomes the boundary of $M^{d+1}$. 
We now turn to the question of under which conditions the resulting manifold $M^{d+1}$ can be considered an asymptotically AdS spacetime, with $\Sigma=\partial M$.  When such conditions are imposed, one may think of the boundary $\Sigma$ as being \emph{reconstructed} from $\Sigma_\star$. 

To begin with, we note that, close to $\Sigma$, the topology of $M^{d+1}$ is the same as the topology of the asymptotic boundary region of an asymptotically AdS space.  Indeed, recalling from~\eqref{X} that $X^{d+2}=D^2/\mathbb Z_n \times\Sigma^d$, we once again observe that the $n\to\infty$ limit corresponds to the zero-radius limit of the boundary circle $S_{r_n}^1=\partial (D^2/\mathbb Z_n)$; see figure~\hyperlink{Fig:1}1. 
It follows that
\begin{equation}
X^{d+2}\rightarrow M^{d+1}:= I\times \Sigma^d \quad{\rm as}\quad n\to\infty~,
\end{equation}
where $I=[0, R_0]$ is the interval that results from the large $n$ limit of the quotient $D^2/\mathbb Z_n$.
Thus,  upon sending $n$ to infinity and zooming into the region close to the origin\footnote{Note that, since we are only focusing on a small region about the origin, whatever happens at the right end of the interval $I$ is irrelevant to us.}, the topology of the space collapses to the collar  $\Sigma \times [0,\varepsilon)$, for some $\varepsilon>0$; this is precisely the cylinder topology of the asymptotic boundary region of an arbitrary AAdS spacetime, in the sense of~\S\ref{sec31}.

On geometric grounds, $M^{d+1}$ will be asymptotically AdS if, locally about $\Sigma=\partial M$, its line element can be written in Graham--Lee normal form;  in our case, this means that the metric
\begin{equation}\label{gM}
g_M = g_X\vert_{n\to\infty}=\frac{R_0^2 \ext\theta^2 + h}{u^2}~,
\end{equation}
which is the metric on $M^{d+1}$ inherited from the metric on $X^{d+2}$ once $n$ is sent to infinity, should equal~\eqref{normal} for $\theta\ll1$. It is direct to verify that this will indeed be the case if we impose
\begin{equation}\label{uRinfty}
u\rightarrow\theta~~{\rm and}~~R_0\to\ell \quad {\rm as}~~n\to\infty~,
\end{equation}
and redefine the radial coordinate as~$r:=R_0 \theta$. 

Consequently and in addition to imposing~\eqref{uRinfty}, for $M^{d+1}=X^{d+2}\vert_{n\to\infty}$ to be asymptotically AdS, the metric~\eqref{gM} should satisfy the \emph{local boundary} equations~\eqref{rr}-\eqref{FGeq}, which thus should arise in the large $n$ limit of the \emph{local approximation} of the \emph{global defect} equations~\eqref{Eett}-\eqref{Eeij}. In other words, the known local equations for the asymptotic AdS boundary should be re-obtained through the following sequence: 
\begin{center}
\begin{tikzpicture}
\node at (-5,0.3){Global defect};
\node at (-5,-0.3){eqs. \eqref{Eett}-\eqref{Eeij}};
\draw[->,line width=0.4mm](-3.2,0)--(-1.8,0);
\node at (-2.5,0.5){$\theta\ll1$};
\node at (0,0.3){Local defect};
\node at (0,-0.3){eqs.};
\draw[->,line width=0.4mm](1.7,0)--(3,0);
\node at (2.4,0.5){$n\to\infty$};
\node at (5,0.3){Local boundary};
\node at (5,-0.3){eqs. \eqref{rr}-\eqref{FGeq}};
\end{tikzpicture}
\end{center}
 
The local defect equations (that complete the center of the diagram above) are given by the $\theta\ll1$ approximation of equations~\eqref{Eett}-\eqref{Eeij}; recalling that at leading order $\cot\theta\approx1/\theta$,  these~are 
\begin{align}
\label{eett}
0&\approx-\frac{1}{2}{\rm Tr}(h^{-1}h'')+\frac{1}{4}{\rm Tr}(h^{-1}h'h^{-1}h')+\frac{1}{2}\frac{u'}{u}{\rm Tr}(h^{-1}h')\cr
&\hspace{6.5cm}+1+\frac{1}{\theta}\frac{u'}{u}+(d+1)\bigg[\frac{u''}{u}-\bigg(\frac{u'}{u}\bigg)^2+\frac{R_0^2}{L^2 u^2}\bigg]~,\\
\label{eeti}
0&\approx h^{jk}(\nabla_j h'_{ik} -\nabla_i h'_{jk})~,\\
\label{eeff}
0&\approx\frac{\theta^2}{n^2}\bigg[1+\frac{u''}{u}+\bigg(\frac{u'}{u}-\frac{1}{\theta}\bigg)\bigg(\frac{1}{2}{\rm Tr}(h^{-1}h')-(d+1)\frac{u'}{u}\bigg)+\frac{(d+1)R_0^2}{L^2u^2}\bigg]~,\\ 
\label{eeij}
0&\approx R_{ij}(h)-\frac{1}{2R_0^2}h''_{ij}  +\frac{1}{2R_0^2} (h'h^{-1}h')_{ij}- \frac{1}{2R_0^2}\bigg(\frac{1}{\theta}-d\,\frac{u'}{u}+\frac{1}{2}{\rm Tr}(h^{-1}h')\bigg) h'_{ij}\cr
&\hspace{4cm}+\frac{1}{R_0^2}\bigg[\frac{u''}{u}+\frac{1}{\theta}\frac{u'}{u} +\frac{d+1}{u^2}\bigg(\frac{R_0^2}{L^2}-(u')^2\bigg)+\frac{1}{2}\frac{u'}{u}{\rm Tr}(h^{-1}h')\bigg] h_{ij}~,
\end{align}
 where thus the symbol ``$\approx$" stands for small $\theta$ approximation. 
 
It is now direct to verify that the boundary equations~\eqref{rr}-\eqref{FGeq} can be obtained from the $n\to\infty$ limit of the defect equations~\eqref{eett}-\eqref{eeij}. 
To this end, we first observe that, since the radial coordinate $r=R_0\theta$, the derivative $\partial_\theta= R_0 \partial_r$, so that each prime in the defect equations differ by a factor of $R_0$ with respect to a prime in the boundary equations. 

We next observe that equation $\eqref{eett}\to\eqref{rr}$ as $n\to\infty$ if, in this limit, the second line in~\eqref{eett} vanishes and $u\to\theta$. The last requirement is part of  condition~\eqref{uRinfty}, and in particular implies that, upon expanding $u\approx u_0+u_1\theta+u_2\theta^2$, the coefficients $u_0\to0$, $u_1\to1$ and $u_2\to0$ in that limit. As for the second line in~\eqref{eett}, using the above expansion for $u$, we can write
\begin{align}
1&+\frac{1}{\theta}\frac{u'}{u}+(d+1)\bigg[\frac{u''}{u}-\bigg(\frac{u'}{u}\bigg)^2+\frac{R_0^2}{L^2 u^2}\bigg] \\
&=\frac{1}{u^2}\bigg\{ \frac{u_1u_0}{\theta}
+\Big[u_0^2 + u_1^2 + (d+1)\big( 2u_0u_2 -u_1^2+ \tfrac{R_0^2}{L^2} \big)\Big]
+\theta\big[  2u_0u_2 +(1-2d)u_1u_2\big] + \mathcal O(\theta^2) \bigg\}\nonumber~.
\end{align}   
Recalling that $u_0\to0$, $u_1\to1$ and $u_2\to0$ as $n\to\infty$, we conclude that the above display vanishes in that limit if we require 
\begin{equation}\label{RL}
\frac{R_0^2}{L^2}\to \frac{d}{d+1} \quad {\rm as}~~n\to\infty~,
\end{equation}
which thus guarantees that equation~\eqref{eett} gives~\eqref{rr} when $n$ is large.

Equation~\eqref{eeti} trivially gives~\eqref{ri} when $n\to\infty$. Also, in this limit, equation~\eqref{eeff} is identically satisfied without imposing any further constraint on the geometry of $M^{d+1}$. Finally and by virtue of~\eqref{uRinfty} and~\eqref{RL}, it is also direct to show that equation~\eqref{FGeq} follows from~\eqref{eeij}. 

Having looked at the $n\to\infty$ regime of the local defect equations, from which we infer that they reproduce the  boundary equations if both conditions~\eqref{uRinfty} and~\eqref{RL} hold, we will next show that they can formally be solved order by order in $\theta$;  in the next subsection, we will explicitly construct this expansion up to second order.

\subsection{Local defect geometry}
\label{sec33}

We now turn to the construction of an asymptotic solution to the local defect equations~\eqref{eett}-\eqref{eeij}---which we recall are valid in the regime where $n>1$ is finite---, of the form
\begin{align}\label{huexp}
h(\theta, x) &= h_{(0)} + \theta h_{(1)}(x) + \theta^2 h_{(2)}(x)+\cdots \cr
u(\theta) &= u_0 + u_1 \theta + u_2 \theta^2 + \cdots
\end{align}
where $h_{(0)}$ is the induced metric on $\Sigma_\star$ and $u_0$ is the defining function zero-mode which we assume to be non-vanishing for finite $n$. 
For simplicity and because it suffices for our purposes, we will consider an expansion up to second order in $\theta$, and leave a more technical analysis of the higher order terms, including possible obstructions, for a separate work.

In order to determine the expansion coefficients in~\eqref{huexp}, we first rewrite equations~\eqref{eett}, \eqref{eeff} and \eqref{eeij} respectively as
\begin{align}
\label{tteq}
0&=\frac{u'}{u} + \theta \mathcal F_{\theta\theta}~,  \\
\label{ffeq}
0&=\frac{u'}{u} -\frac{1}{2(d+1)}{\rm Tr}(h^{-1}h') + \theta\mathcal F_{\phi\phi}~, \\
\label{ijeq}
0&=\frac{u'}{u}\,h_{ij} -\frac12\,h'_{ij} +  \theta\mathcal F_{ij}~, 
\end{align}
where 
\begin{align}
\mathcal F_{\theta\theta} &:=-\frac{1}{2}{\rm Tr}(h^{-1}h'')+\frac{1}{4}{\rm Tr}(h^{-1}h'h^{-1}h')+\frac{1}{2}\frac{u'}{u}{\rm Tr}(h^{-1}h')\cr
&\hspace{7cm}+1+(d+1)\bigg[\frac{u''}{u}-\bigg(\frac{u'}{u}\bigg)^2+\frac{R_0^2}{L^2 u^2}\bigg]~,  \\
\mathcal F_{\phi\phi} &:=\frac{1}{d+1}\bigg(1+\frac{u''}{u}+\frac{1}{2}\frac{u'}{u}{\rm Tr}(h^{-1}h')\bigg) +\frac{1}{u^2}\bigg(\frac{R_0^2}{L^2}-(u')^2\bigg)~, \\
\mathcal F_{ij}&:= R_0^2 R_{ij}(h) -\frac{1}{2}h''_{ij}  +\frac{1}{2}(h'h^{-1}h')_{ij}+ \frac{1}{2}\bigg(d\,\frac{u'}{u}-\frac{1}{2}{\rm Tr}(h^{-1}h')\bigg) h'_{ij}\cr
&\hspace{5cm}+\bigg[\frac{u''}{u} +\frac{d+1}{u^2}\bigg(\frac{R_0^2}{L^2}-(u')^2\bigg)+\frac{1}{2}\frac{u'}{u}{\rm Tr}(h^{-1}h')\bigg] h_{ij}~.
\end{align}
As in the case of the boundary equation~\eqref{ri}, the defect equation~\eqref{eeti} will not play any role due to one of the Bianchi identities, and we have thus not considered it in the above two displays. 
Importantly, we are a priori assuming that all the $\mathcal F$'s defined above are finite at $\theta=0$ and hence ($\theta \mathcal F)\vert_{\theta=0}=0$. We will a posteriori realize that this is not really an assumption but a consistency condition. 

The first order expansion coefficients in the anstaz~\eqref{huexp} can be determined by first evaluating~\eqref{tteq} at $\theta=0$. This implies $u'\vert_{\theta=0}=0$, so that the defining function $u$ can not have a linear term at finite $n$ (but because of the condition~\eqref{uRinfty}, note that $u$ will actually be linear for large~$n$). 
Using this fact and further evaluating~\eqref{ffeq} or~\eqref{ijeq} at $\theta=0$, it follows that, since $h_{(0)}$ is the induced metric on $\Sigma_\star$ and thus it is non-degenerate, the linear term $h'\vert_{\theta=0}=0$.  Then, at this order we conclude that
\begin{equation}\label{linear}
u_1 =0\,\quad {\rm and} \quad h_{(1)}=0\,.
\end{equation}

The second-order coefficients follow from the first derivative of the defect equations. Taking the derivative of~\eqref{tteq} and evaluating at the origin using~\eqref{linear} gives
\begin{equation}
0= \bigg (\frac{u''}{u} + \mathcal F_{\theta\theta} \bigg)\bigg\vert_{\theta=0}
=2\frac{u_2}{u_0}- {\rm Tr} (h_{(0)}^{-1}h_{(2)})+1 +2(d+1)\frac{u_2}{u_0} + \frac{(d+1)R_0^2}{L^2u_0^2}~,
\end{equation}
where the last four terms come from $\mathcal F_{\theta\theta}\vert_{\theta=0}$. It follows that 
\begin{equation}\label{traceh2}
{\rm Tr}(h^{-1}_{(0)} h_{(2)}) = 1+2(d+2) \frac{u_2}{u_0} +  \frac{(d+1)R_0^2}{L^2u_0^2}~.
\end{equation}  
It is not hard to check that the derivative of~\eqref{ffeq} gives the same information as~\eqref{traceh2}.  As for the derivative of~\eqref{ijeq}, recalling that $h'\vert_{\theta=0}=0$ and $u'\vert_{\theta=0}=0$, we have that
\begin{equation}
0=\bigg(\frac{u''}{u}\,h_{ij} -\frac12\,h''_{ij} + \mathcal F_{ij} \bigg)\bigg\vert_{\theta=0}
= 2\frac{u_2}{u_0}\,h^{(0)}_{ij} - h^{(2)}_{ij} + R_0^2 R^{\Sigma_\star}_{ij} - h^{(2)}_{ij} + \bigg[2\frac{u_2}{u_0} +\frac{ (d+1) R_0^2}{L^2u_0^2}\bigg] h^{(0)}_{ij}~,
\end{equation}
where the last four terms follow from $\mathcal F_{ij}\vert_{\theta=0}$, and where we have denoted by $R^{\Sigma_\star}_{ij}=R_{ij}(h)\vert_{\theta=0}$ to the Ricci tensor of the defect submanifold $(\Sigma_\star, h_{(0)})$. Thus, solving for the second-order coefficient gives
\begin{equation}\label{fullh2}
h^{(2)}_{ij} = \frac{R_0^2}{2} R^{\Sigma_\star}_{ij} + \bigg[2\frac{u_2}{u_0} + \frac{(d+1)R_0^2}{2L^2u_0^2} \bigg] h^{(0)}_{ij}~,
\end{equation}
whose trace should then be consistent with~\eqref{traceh2}; this fixes the curvature of the defect in terms of the defining function expansion coefficients $u_0$ and $u_2$, and the scales $R_0$ and $L$
\begin{equation}\label{Riccistar}
R_{\Sigma_\star} = \frac{2}{R_0^2} \bigg[1+ 4\frac{u_2}{u_0} + \bigg(1-\frac{d}{2}\bigg) \frac{(d+1)R_0^2}{L^2u_0^2} \bigg]~.
\end{equation}
The above implies that the defect submanifold is constrained to have constant curvature.

Because it will be useful afterwards, let us write down the explicit form of the above solution for the case in which $\Sigma_\star$ is a 2-manifold. Note that, when $d=2$, the last term in~\eqref{Riccistar} vanishes identically, so that the defect curvature is simply given by
\begin{equation}\label{Rstar2}
R_{\Sigma_\star} = \frac{2}{R_0^2}\bigg(1+4\,\frac{u_2}{u_0}\bigg) ~.
\end{equation}
When the above is the case, the defining function
\begin{equation}
u = u_0 \bigg[ 1 + \frac{1}{4}\bigg(\frac{R_0^2}{2} R_{\Sigma_\star}-1\bigg) \theta^2\bigg]~,
\end{equation}
and the second-order expansion coefficient
\begin{equation}
h^{(2)}_{ij} = \frac{R_0^2}{2} R^{\Sigma_\star}_{ij} + \bigg(2\frac{u_2}{u_0} + \frac{3R_0^2}{2L^2u_0^2} \bigg) h^{(0)}_{ij}~.
\end{equation} 
Equation~\eqref{Rstar2} will be of particular relevance in~\S\ref{sec42}.

\paragraph{A consistency check.}
The local formul\ae~\eqref{linear}, \eqref{fullh2}-\eqref{Riccistar} can be scrutinized by studying how the global family of solutions~\eqref{Fh} spelled out in \S\ref{sec21} behaves close to the location of the defect~$(\Sigma_\star, h_{(0)})$. 
In the global case, we have that
\begin{align}\label{testsol}
h(\theta, x) = h_{(0)}(x)~, \quad
u(\theta) =\cos\theta~, \quad R_0=L~,
\end{align}
where $h_{(0)}$ satisfy the Einstein constraint~\eqref{Econst}, which in turn implies
\begin{equation}\label{Riccitest}
R_{\Sigma_{\star}}=-\frac{d(d-1)}{L^2}~,
\end{equation}
with $R_{\Sigma_{\star}}=R(h_{(0)})=R(h)\vert_{\theta=0}$.

Close to the defect, the defining function in~\eqref{testsol} goes as $u\approx1-\frac12\theta^2$,  so that the expansion coefficients $u_0=1$ and $u_2=-1/2$; interestingly, substituting these values in the local formula~\eqref{Riccistar} (remembering from~\eqref{testsol} that $R_0=L$) we obtain \emph{exactly}~\eqref{Riccitest}.

It is also interesting to note that the global solution for $h$ in~\eqref{testsol} has vanishing second-order coefficient $h_{(2)}$; replacing $u_0=1$, $u_2=-1/2$ and $R_0=L$ in the local formula~\eqref{fullh2} and imposing $h_{(2)}=0$ gives now \emph{exactly} the Einstein condition~\eqref{Econst}.  In other words, we have learnt that the Einstein constraint on $h_{(0)}$ is equivalent to the vanishing of $h_{(2)}$.

\paragraph{Defining function as an order parameter.}

It is important to note that condition~\eqref{uRinfty} and solution~\eqref{linear}, \eqref{fullh2}-\eqref{Riccistar} imply that the defining function $u$ exhibits rather different behaviors depending on whether the submanifold $\Sigma$ is in a defect phase (finite $n$) or a boundary phase (large $n$).  Indeed, condition~\eqref{uRinfty} enforces the defining function to become linear in the $n\to\infty$ limit, suppressing the zero-mode $u_0$ and second-order coefficient $u_2$ in the boundary phase. Solution~\eqref{linear}, on the other hand, requires no contribution from the linear term in $u$-expansion within the defect phase.  This behavior suggests the existence of a (presumably abrupt) phase transition, whereby the defining function may be thought of as the order parameter;  the situation is qualitatively represented down below.

\vskip1cm
\begin{center}
\begin{tikzpicture}[scale=0.8]
\begin{axis}[thick, xmin = 0, xmax = 5,ymin = 0, ymax = 10,axis lines* = left, xtick = \empty, ytick = \empty, clip = false] 
\addplot[dashed, domain = 0:5, restrict y to domain = 0:10, samples=100, line width=0.5mm, color = darkred]{6*(1+(1/4)*x^2)};
\addplot[dashed,domain = 0:5, restrict y to domain = 0:10, samples=100, line width=0.5mm, color = darkred]{5*(1+(1/4)*x^2)};
\addplot[dashed,domain = 0:5, restrict y to domain = 0:10, samples=100, line width=0.5mm, color = darkred]{4*(1+(1/4)*x^2)};
\addplot[dashed,domain = 0:5, restrict y to domain = 0:10, samples=100, line width=0.5mm, color = darkred]{3*(1+(1/4)*x^2)};
\addplot[dashed,domain = 0:5, restrict y to domain = 0:10, samples=100, line width=0.5mm, color = darkred]{2*(1+(1/4)*x^2)};
\addplot[dashed,domain = 0:5, restrict y to domain = 0:10, samples=100, line width=0.5mm, color = darkred]{1*(1+(1/4)*x^2)};
\addplot[domain = 0:5, restrict y to domain = 0:10, line width=0.4mm, color = darkblue]{x};
\end{axis}
\draw [thick,->] (0,0) --(0,6);
\draw [thick,->] (0,0) --(7,0);  
\node at (0,6.5) {$u\approx u_0+u_1\theta+u_2\theta^2$};
\node at (7.3,0) {$\theta$};
\node [darkred] at (7,6) {\small $T>0$};
\node [darkred] at (7,5.4) {\small defect phase};
\node [darkred] at (7,4.8) {\small ($u_1=0$)};
\node [darkblue] at (5.6,1.7){\small $T=0$};
\node [darkblue] at (5.6,1.1) {\small boundary phase};
\node [darkblue] at (5.6,0.5) {\small ($u_0=0=u_2$)};
\node[text width=15cm, text justified] at (3,-3){{\hypertarget{Fig:3}\bf Fig.~3}: 
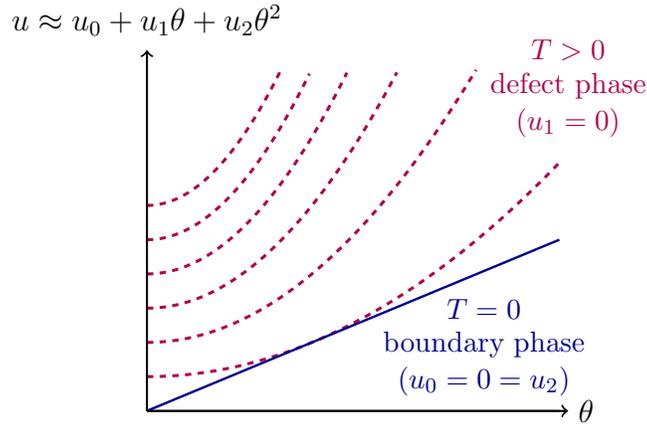
Asymptotic behavior of the defining function $u$. The defect phase is defined by a finite temperature $T=1/n$ and the absence of linear terms in $u$;  all the dashed red curves depicted have equal $u_2/u_0=1/4$. The boundary phase is reached at zero temperature, in which case the defining function becomes linear with no zero-mode. In blue we display the $u_1=1$ defining function.};
\end{tikzpicture}
\end{center}

\section{Defect central charge and its boundary limit}
\label{sec4}
In \S\ref{sec31} we argued that the zero-radius limit $n\to\infty$ defines a defect-to-boundary transition, in which the defect submanifold $\Sigma^d_\star\hookrightarrow X^{d+2}$ becomes the boundary $\Sigma^d=\partial M^{d+1}$ of an asymptotically AdS space $M^{d+1}$. 
Motivated by this transition, in this section we argue that it is plausible to think that, at least in some non-trivial cases, the holographic CFT at the asymptotic boundary of AdS$^{d+1}$ is truly a reincarnation of some defect, infrared field theory with support on the interior of $X^{d+2}$.
Here, we collect some arguments supporting the existence of such a theory.  

\subsection{Heuristics}
\label{sec41}
\paragraph{Symmetries.}  There is a simple symmetry argument, similar to the one used to justify the existence of a CFT at the boundary of AdS, which can be equally invoked to support the existence of a CFT on $\Sigma_\star$. To this end, recall that the symmetry group of AdS$^{d+2}$ has as a subgroup
\begin{equation}\label{isometries}
SO(2, d+1)\supset SO(p) \times SO(2,q)~,\quad p+q=d+1~.
\end{equation}
In our construction, because $X^{d+2}=D^2/\mathbb Z_n\times \Sigma^d$ and the explicit form of the metric~\eqref{gX} and~\eqref{gXcoords}, each factor at the right-hand side of~\eqref{isometries} correspond to the (manifest) isometries of one factor in the decomposition of $X^{d+2}$;  $SO(p)$ corresponds to the symmetries of the quotient $D^2/\mathbb Z_n$ and $SO(2,q)$ to the symmetries of $\Sigma^d$. Away from the singular point $\theta=0$, the symmetries of the former are those of a circle, which fix $p=2$ and consequently $q=d-1$, so that $\Sigma^d$ acquires~$SO(2,d-1)$ symmetries. But at the singular point, the circle above shrinks to a point, which fix $p=1$ and in turn equips the defect submanifold $\Sigma^d_\star$ with $SO(2,d)$ symmetry. This is of course the full conformal group in dimension $d$.  
\bigskip
\begin{center}
\begin{tikzpicture}
\draw (-1.35,0) -- (0,2);
\draw (1.35,0) -- (0,2);
\node [darkred] at (0,2) {\textbullet};
{\color{darkblue}
\draw[thick, dashed] (1,0.5) 
arc[start angle=0,end angle=180, 
x radius=1, y radius=0.2];
\draw[thick] (1,0.5) 
arc[start angle=0,end angle=-180, 
x radius=1, y radius=0.2];
}
\node at (0,-0.5) {$D^2/\mathbb Z_q\times\Sigma^d$};
\node at (4,2) {$\{p\}\times\Sigma_\star^d\cong SO(1)\times SO(2,d)$};
\draw[thick,->, darkred](1.2,2)--(0.3,2);
\node at (4.5,0.5) {$S^1\times\Sigma^d\cong SO(2)\times SO(2,d-1)$};
\draw[thick,->, darkblue](1.5,0.5)--(1.1,0.5);
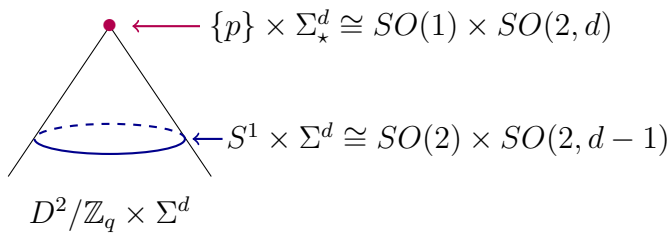
\node[text width=14cm, text justified] at (3,-2.2){{\hypertarget{Fig:4}\bf Fig.~4}: The symmetries of the manifold $X^{d+2}$. Away from the center of the disk, $\Sigma^d$ has $SO(2,d-1)$ symmetry. At the locus of the orbifold singularity, these symmetries are enhanced to those of the conformal group $SO(2,d)$.
};
\end{tikzpicture}
\end{center}

\paragraph{Existence of a local stress tensor.} 
Strictly speaking, in AdS space, any submanifold at some fixed radius has as a symmetry group the conformal group in one lower dimension. However, among all these codimension-one submanifolds, there exists one and only one equipped with a stress-energy tensor; this is the AdS boundary, endowed with the Brown--York stress tensor~\cite{Brown:1992br, Balasubramanian:1999re}. 

From the geometric point of view, the existence of a local stress-energy tensor on a given submanifold is related to the way this submanifold is embedded into the full space. In the case of the boundary $\Sigma^d$ of an asymptotically AdS space $M^{d+1}$,  there exists a gauge (the Graham--Lee normal form~\eqref{normal}) in which the spacetime metric blows up at the location of the boundary. This means that the boundary submanifold $\Sigma^d$ can be thought of as being embedded into $M^{d+1}$ via a one-dimensional delta function\footnote{Another way to see this is by gluing two copies of $M$ along their conformal boundary. The gluing procedure enhances a $\mathbb Z_2$-symmetry whereby the two fully overlapped conformal infinities become a single domain wall. Because of the latter symmetry, the metric will contain the absolute value of the radial coordinate whose second derivative is a delta function (in codimension one).}
\begin{equation}\label{bdryemb}
\Sigma^d\stackrel{\delta}\hookrightarrow M^{d+1}\quad (\Sigma=\partial M). 
\end{equation}
It is precisely the existence of such a singular embedding which permits the insertion of a stress-energy operator at the location of the boundary, and such a stress tensor in principle suffices to define a CFT\footnote{From an axiomatic point of view, a stress tensor is a sufficient but not necessary condition to define a conformal field theory. Indeed, there exist a number of CFTs that have no stress energy tensor~\cite{Rychkov:2016iqz}.}.

A similar reasoning in one higher codimension applies to a $\mathbb Z_n$-orbifold defect. In this case, when $n>1$ is finite, the codimension-two set of fixed points~$\Sigma^d_\star$ embeds into $X^{d+2}=D^2/\mathbb Z_n\times \Sigma^d$ via a delta function in codimension two, that is
\begin{equation}\label{defemb}
\Sigma^d_\star\stackrel{\delta^2}\hookrightarrow X^{d+2} \quad (\Sigma_\star={\rm defect},~{\rm finite}~n>1).
\end{equation}
This is because the $\mathbb Z_{n>1}$ action locally induces a conically singular geometry about the center of the disk and, on this background, some of the components of the Einstein tensor contain a term of the form~\cite{Fursaev:1995ef} (here we take $\rho=R_0\theta$ so that the metric about the center of $D^2/\mathbb Z_n$ is locally given by $\ext\rho^2 + n^{-2}\rho^2\ext\phi^2$)
\begin{equation}\label{singterm}
\bigg(1-\frac{1}{n}\bigg)\nabla^2\log\rho\sim\bigg(1-\frac{1}{n}\bigg)\delta^2(\rho)~,
\end{equation}
which is not present in the smooth case $n=1$. Due to~\eqref{singterm} and in order to have a well defined variational principle, one needs to couple to the gravitational action a Nambu--Goto term with support on $\Sigma_\star$, which in turns fixes the form of the stress-energy tensor to 
\begin{equation}\label{T}
T^{\Sigma_\star}_{ij} = \frac{1}{4G_{d+2}}\bigg(1-\frac1n\bigg) h^{(0)}_{ij}~,
\end{equation}
where we recall that $h^{(0)}_{ij}=h_{ij}(0, x)$ denotes the induced metric on $\Sigma_\star$. Hence, just as in the boundary case, the existence of~\eqref{T}---whose insertion is possible because of the singular embedding~\eqref{defemb} and whose precise form is determined by consistency of the variational principle---is indicative of the existence of a CFT on $\Sigma_\star$. 
 
\paragraph{Defect conformal structure.}  The common lore states that the boundary of AdS is special because it carries a conformal structure. Although this is true, conformal structures can in general be attached to \emph{any} submanifold embedded in AdS. 

A conformal structure is a metric-dependent\footnote{The boundary of a manifold, on the other hand, is a metric-independent notion that only depends on the topology of the manifold, regardless of which metric one puts on it.} notion which refers to an equivalence class of metrics on a given submanifold.  
Consider for instance the conformal infinity of $(X^{d+2}, g_X)$ (as defined in~\S\ref{sec2}), whose location coincides with that of the boundary of $X$ at $\theta=\pi/2$. The induced metric on this submanifold is
\begin{equation}\label{hSX}
g_{\partial X} = u^2\,g_X\Big\vert_{\theta=\frac{\pi}{2}}=\frac{R_0^2}{n^2} \ext\phi^2 + h\big(\tfrac{\pi}{2}, x\big)~.
\end{equation}
Then, the fact that the defining function $u$ is not unique implies that the rescaling
\begin{equation}\label{omegau}
u\rightarrow \Omega u
\end{equation}
(where $\Omega$ is a positive smooth function with no poles at $\theta=\pi/2$) induces the conformal class of metrics
\begin{equation}\label{hconfinf}
\big[g\big]_{\partial X}=\Omega^2\big(\tfrac{\pi}{2}\big) g_{\partial X}~,
\end{equation}
on the boundary of $X$.

The same argument above applies to the defect submanifold $\Sigma_\star=X\vert_{\theta=0}$. Indeed, due to the non-uniqueness of the defining function~\eqref{omegau}, $\Sigma_\star$ is naturally equipped with the conformal class
\begin{equation}\label{hdef}
\big[h\big]_{\Sigma_\star}=\Omega^2(0) h(0,x)~, 
\end{equation}
where $h(0,x)=h_{(0)}$ is the induced metric on $\Sigma_\star$. Note that, because of the boundary reconstruction discussed in \S\ref{sec32}, the conformal structure~\eqref{hdef} becomes the conformal structure at the boundary of $M^{d+1}$ when $n\to\infty$.

\subsection{Defect central charge and its boundary limit}
\label{sec42}
Because of the arguments given in~\S\ref{sec41}, we hereafter assume the existence of a conformal field theory on~$\Sigma^d_\star$. 

Our aim is now to illustrate with a simple example how the $\Sigma_\star\to\Sigma$ transition amounts to computing the central charge of the holographic boundary CFT, defined on $\Sigma^d=\partial M^{d+1}$, from the central charge of the CFT defined on $\Sigma^d_\star$. 

To this end, we specialize to the $d=2$ case and consider a two-dimensional defect embedded in a 4-manifold.  As discussed in \S\ref{sec41}, when $n>1$, the singular embedding $\Sigma^2_\star\hookrightarrow X^4$ amounts to the insertion of the local stress-energy tensor~\eqref{T} with support on $\Sigma_\star$, which we recall is given by
\begin{equation}\label{T2}
T^{\Sigma_\star}_{ij} = \frac{1}{4G_4}\bigg(1-\frac1n\bigg) h^{(0)}_{ij}~,
\end{equation}
where $h^{(0)}_{ij}$ is the induced metric on $\Sigma_\star$ and $G_4$ is Newton constant.

\paragraph{Defect central charge.}
From the defect point of view, the trace of~\eqref{T2} is classically anomalous in the sense that
\begin{equation}\label{Tran}
{\rm Tr}(h^{-1}_{(0)}T_{\Sigma_\star})= \frac{c_\star}{24\pi}\, R_{\Sigma_\star}~,
\end{equation}
where $c_{\star}$ denotes the central charge of the CFT on $\Sigma^2_{\star}$.  
In the above, since ${\rm Tr}(h^{-1}_{(0)}h_{(0)})=2$, the left-hand side gives
\begin{equation}\label{Tran2}
{\rm Tr}(h^{-1}_{(0)}T_{\Sigma_\star})= \frac{1}{2G_4}\bigg(1-\frac1n\bigg)~,
\end{equation}
while the defect Ricci scalar at the right-hand side was determined in~\eqref{Rstar2}; it crucially depends on the ratio of the defining function expansion coefficients, that we denote by $\mu$:
\begin{equation}\label{Rstar22}
R_{\Sigma_\star} = \frac{2}{R_0^2}(1+4\mu)~, \quad \mu:=\frac{u_2}{u_0} ~.
\end{equation}
From~\eqref{Tran}, \eqref{Tran2} and \eqref{Rstar22}, and recalling that Newton constant can be dimensionally reduced \emph{\`a la} Kaluza--Klein as $G_4={\rm Vol} (S^1_{R_0}) G_3=2\pi R_0 G_3$ (we denote by ${\rm Vol}(S^1_{R_0})$ the volume of the transverse circle of radius $R_0$), it follows that
\begin{equation}\label{cstar}
c_\star = \bigg(1-\frac1n\bigg) \frac{3R_0}{(1+4\mu) G_3}~.
\end{equation}

Equation~\eqref{cstar} provides a formal expression for the central charge $c_\star$ of the field theory on $\Sigma_\star$ in terms of the curvature of the defect---which is in turn controlled by $\mu:= u_2/u_0$---, the scale $R_0$, and Newton constant in dimension three.  

Note that, since orbifold parameter $n>1$ and the radius $R_0>0$, the sign of $c_\star$ is controlled by the curvature coefficient $\mu$. This means that unitarity of the theory on $\Sigma_\star$ depends on the curvature of that manifold; defects whose curvature $4\mu>-1$ will support unitary theories, while defects with $4\mu<-1$ will admit non-unitary ones. 

\paragraph{Boundary limit.} 
Let us conclude by thinking of the $\Sigma_\star\to\Sigma$ transition. Recalling from~\eqref{uRinfty} that, when $n$ is large, $R_0\to\ell$ (where $\ell$ is the AdS$^{d+1}$ radius), it follows that
\begin{equation}\label{cbdry}
c_\star\to c=  \frac{3\ell}{(1+4\mu) G_3} \quad {\rm as} \quad n\to\infty\,.
\end{equation}
Thus, the resulting boundary central charge, denoted by $c$, will necessarily retain the information about the curvature of the parent defect $\Sigma_\star$ from which the boundary submanifold $\Sigma$ emerge in the limit $n\to\infty$. 

Importantly, because of the $\mu$-dependence of~\eqref{cbdry}, we observe that only a defect with positive, $\mu=1/4$ curvature\footnote{Note that this is consistent with the fact that stability of the dual CFT requires a positively curved boundary~\cite{Seiberg:1999xz}.} will give rise to a boundary CFT with central charge equal to the Brown--Henneaux~\cite{Brown:1986nw} central charge $c=3\ell/2G_3$. 
Indeed, according to our findings, there exist a number of theories on $\Sigma_\star$ whose boundary limit gives rise to holographic, possibly non-unitary theories with different values of their central charge. 
For instance, for the background constructed in~\S\ref{sec22}, the defining function $u=\cos\theta\approx 1-\frac12 \theta^2+\cdots$, so that the expansion coefficients $u_0=1$ and $u_2=-1/2$. In that case $\mu=-1/2$ and thus the resulting boundary theory is a non-unitary CFT with central charge $c=-3\ell/G_3$. 
A diagram with the space of possible defect theories and their boundary limit is depicted in figure~\hyperlink{Fig:5}5 below.

\bigskip
\begin{center}
\begin{tikzpicture}[scale=0.9]
\fill[fill=yellow!6] (0,-2)--(0,4)--(5,4)--(5,-2);
\fill[fill=red!6] (0,-2)--(0,-4)--(5,-4)--(5,-2);
\node at (2.5,-0.7) {Unitary CFT's};
\node at (2.5,-1.2) {$c_\star>1$};
\draw [line width=0.4mm,->] (4.2,-1.2)--(4.2,-0.4); 
\node at (2.6,-2.8) {Non-unitary CFT's};
\node at (2.6,-3.3) {$c_\star<1$};
\draw [line width=0.4mm,->] (0.5,-2.5)--(0.5,-3.3); 
\draw [line width=0.5mm,dashed,darkblue] (0,2) --(5,2); 
\draw [line width=0.5mm,dashed,darkred] (0,-2) --(5,-2); 
\draw [darkblue, line width=0.6mm,->] (0,-4.5) --(0,4.5); 
\node at (0,5) {$\mu=u_2/u_0$};
\draw [thick,dashed] (5,-4.5) --(5,4.5);
\node at (5,-5) {$T=1$};
\node at (0,5) {$\mu=u_2/u_0$};
\draw [thick,->] (0,0) --(6,0); 
\node at (7,0) {$T=1/n$};
\node at (0,-5.6) {Boundary CFT's};
\node at (0,-5) {$T=0$};
\node [darkblue] at (0,2) {\LARGE\textbullet};
\node [darkblue] at (6.2,2) {$\mu=+1/4$};
\node [darkred] at (6.2,-2) {$\mu=-1/4$};
\node at (-0.7,0) {$\mu=0$};
\draw [line width=0.4mm,->] (-1,2)--(-0.2,2); 
\node at (-2.5,2.5) {Boundary CFT };
\node at (-2.5,1.9) {with \large$c=\frac{3\ell}{2G_3}$};
%
\node[text width=16cm, text justified] at (2.5,-9.5){
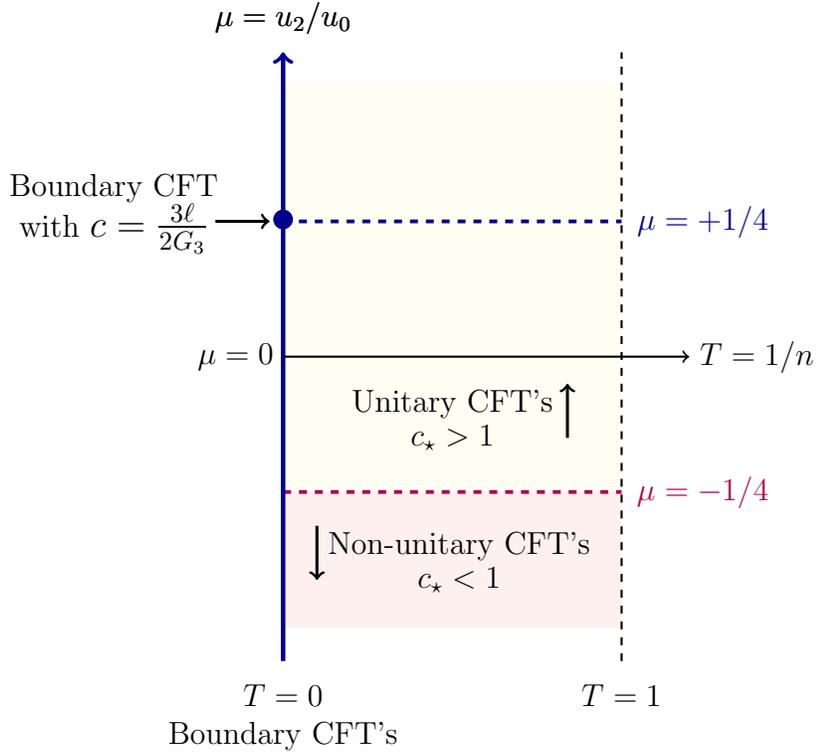
\small {\hypertarget{Fig:5} \bf Fig.~5}: Possible CFTs in $(\mu, T)$ space. The curvature parameter $\mu:=u_2/u_0$ controls the sign of the central charge of a given theory, and the ``temperature"  $T=1/n$ defines two different phases.
For $T>0$ (defect phase), the possible CFTs on $\Sigma_\star$ are unitary for $\mu>-1/4$ and non-unitary otherwise. 
When $n$ is large one reaches the zero-temperature, boundary phase. In this phase, there is a unique point representing a holographic theory with Brown--Henneaux central charge.
Note that, although both $u_0, u_2\to0$ as $n\to\infty$, the ratio $\mu=u_2/u_0$ remains finite.
};
\end{tikzpicture}
\end{center}

\section{Discussion}
\label{discussion}
In this work we have argued that the dynamics and geometry of the boundary of an asymptotically AdS space can be reconstructed from a conical bulk defect embedded in one higher (co)dimension. Consequently, all the properties of the boundary submanifold---including the capability of encapsulating localizable degrees of freedom in an holographic fashion---can be thought of as inherited from a parent bulk defect. 
Based on this idea, we have conjectured that the holographic theory at the boundary of AdS arises in a certain zero-radius limit of a field theory on an infrared defect. 
In order to illustrate our conjecture, we worked out the lowest dimensional case and showed that the Brown--Henneaux central charge arises from the zero-radius limit $r_n\sim1/n\to0$ of the central charge on a two-dimensional defect (at fixed curvature) embedded in four dimensions.

Our findings seem to manifest the need for the inclusion of bulk defects into the holographic framework. Indeed, following the ideas of the holographic renormalization group flow, one may hypothesize that the conformal field theory on the defect $\Sigma_\star$ represents the infrared fixed point of the dual flow, with the boundary dual theory being the ultaviolet fixed point. 
In these regards, the $\Sigma_\star\to\Sigma$ transition here proposed would represent a second type of flow in the space of holographic theories, in which the direction of the flow is reversed with respect to the direction of the renormalization group flow, from the infrared to the ultraviolet, at the cost of suppressing one spacetime dimension.
The situation is sketched below:

\vskip20pt
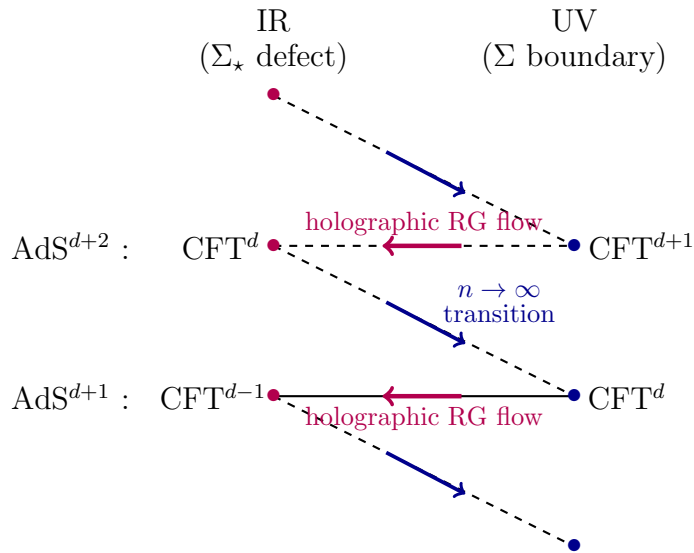
\begin{center}
\begin{tikzpicture}
\node at (-2,3) {IR};
\node at (-2,2.5) {($\Sigma_\star$ defect)};
\node at (2,3) {UV};
\node at (2,2.5) {($\Sigma$ boundary)};
\draw[thick, dashed] (-2,2)--(2,0);
\draw [darkblue, line width=0.5mm,->] (-0.5,1.25) --(0.55,0.7); 
\node [darkred] at (-2,2) {\textbullet};
\draw[thick, dashed] (-2,0)--(2,-2);
\draw [darkblue, line width=0.5mm,->] (-0.5,-0.75) --(0.55,-1.3); 
\node at (1,-0.9) {\footnotesize \color{darkblue} transition};
\node at (1,-0.6) {\footnotesize \color{darkblue} $n\to\infty$};
\draw[thick, dashed] (-2,-2)--(2,-4);
\draw [darkblue,line width=0.5mm,->] (-0.5,-2.75) --(0.55,-3.3); 
\node [darkblue] at (2,-4) {\textbullet};
\draw[thick, dashed] (-2,0)--(2,0);
\node at (-4.7,0) {AdS$^{d+2}$~:};
\node at (-2.7,0) {CFT$^d$};
\node at (2.9,0) {CFT$^{d+1}$};
\node [darkred] at (-2,0) {\textbullet};
\node [darkblue] at (2,0) {\textbullet};
\node at (0,0.3) {\color{darkred}\footnotesize holographic RG flow};
\draw [darkred, line width=0.6mm,<-] (-0.55,0) --(0.5,0);
\draw[thick] (-2,-2)--(2,-2);
\node at (-4.7,-2) {AdS$^{d+1}$~:};
\node at (-2.8,-2) {CFT$^{d-1}$};
\node at (2.7,-2) {CFT$^{d}$};
\node [darkred] at (-2,-2) {\textbullet};
\node [darkblue] at (2,-2) {\textbullet};
\node at (0,-2.3) {\color{darkred}\footnotesize holographic RG flow};
\draw [darkred, line width=0.6mm,<-] (-0.55,-2) --(0.5,-2); 
\end{tikzpicture}
\end{center}

The above elements lead us to speculate that gauge/gravity duality belongs to broader scheme, in which dual gauge theories do not necessarily have support on boundary submanifolds. In such a scheme, bulk gravitational theories ought to be formulated on manifolds with multiple boundaries and extended objects in all possible codimensions; Hilbert spaces are assigned to boundaries---encapsulating states of a boundary, \emph{large} $N$ gauge theory---, as well as to defects---encoding defect states labeled by the codimension number and described by means of a presumable \emph{finite} $N$ gauge theory---. 
Furthermore, Hilbert spaces assigned to boundaries and defects are expected to be related via a (co)dimensional ladder of dualities involving different limits of the moduli parameters of the theory.

Clearly, many open questions remain to be investigated. The very existence of the infrared type of theories postulated in this work, as well as the universality and robustness of our framework require to be further studied; this is the subject matter of some of our current, ongoing research.

\section*{{\large Acknowledgements}}
I am indebted to A.~Waldron for several discussions related to the topic of this article, and his encouragement, support, and coffee invites during my years in Davis. 
I would also like thank to G.~Arenas, F.~Diaz and P.~Sundell for collaboration at the initial stage of this project. 

The main part of this work was partially supported by the fellowship~{\sc Postdoctorado en el Extranjero Becas Chile} N$^{\rm o}$ 74200106, carried out at UC Davis.
I'm currently supported by the grant {\sc Fondecyt Postdoctorado} N$^{\rm o}$ 3220236, hosted by UC Chile. 
I am also grateful to Y.~Burak and the Hebrew University of Jerusalem for the kind hospitality and financial support during the completion of this manuscript.
\appendix
\section{Conventions}
\label{appA}
Through the body of this article, we take the bulk dimension to be
\begin{equation}
D=d+2~,\quad d\geq2~,
\end{equation}
and often indicate the dimension of a manifold as a superscript; we write $X^{d+2}$ to denote a smooth Riemannian manifold of dimension $D=d+2$.  On tensors, we sometimes attach a manifold as a sub or superscript. For instance, we may write $R_{\Sigma}$ to indicate that such a tensor is intrinsically defined on $\Sigma$ or constructed from the induced metric on that manifold.
We omit decorations when all is clear from context. 

Given a metric $g$ compatible with a (Levi-Civita) connection $\nabla g=0$, the Christoffel symbols are given by
\begin{equation}
\Gamma^\rho_{\mu\nu} = \frac{1}{2} g^{\rho\sigma} \big(\partial_\mu g_{\nu\sigma}+\partial_\nu g_{\mu\sigma} - \partial_\sigma g_{\mu\nu} \big)~.
\end{equation}
The components of the Riemann and Ricci tensors, and the Ricci scalar are defined as
\begin{equation}
R_{\mu\nu\rho}{}^{\sigma}= -2\,\partial_{[\mu} \Gamma_{\nu]\rho}^{\sigma} -2\,\Gamma_{\lambda[\mu}^{\sigma}\Gamma_{\nu]\rho}^{\lambda}~, \qquad 
R_{\mu\nu}=R_{\mu\lambda\nu}{}^{\lambda}~,\qquad 
R=g^{\mu\nu} R_{\mu\nu}~.
\end{equation}
Einstein equations are
\begin{equation}\label{EEs}
R_{\mu\nu}-\frac{1}{2} R g_{\mu\nu}+\Lambda g_{\mu\nu}=0~,\qquad 
\Lambda=-\frac{(D-1)(D-2)}{2L^2}<0~,
\end{equation}
or equivalently
\begin{equation}\label{Einstein}
R_{\mu\nu}+ \frac{D-1}{L^2}\,g_{\mu\nu}=0~,
\end{equation} 
where $L$ is the AdS$^{d+2}$ radius. 
\section{Curvatures}
\label{appB}
Here we collect the Christoffel symbols and components of the Ricci tensor involved in the calculation of Einstein equations of \S\ref{sec21}.

Consider the globally defined metric~\eqref{gXcoords} 
\begin{equation}
g_X = \frac{R_0^2(\ext \theta^2 + n^{-2}\sin^2\theta\,\ext\phi^2)+h_{ij}(\theta, x)\ext x^i\ext x^j}{u^2(\theta)}~.
\end{equation}
The non-vanishing Christoffel symbols are
\begin{align}\label{Ggn}
\Gamma^{\theta}_{\theta\theta}=-\frac{u'}{u}~,\quad
\Gamma^{\theta}_{\phi\phi}&= \frac{\sin^2\theta}{n^2}\Big(\frac{u'}{u}-\cot\theta\Big)~,\quad
\Gamma^{\theta}_{ij} =\frac{1}{R_0^2}\Big(\frac{u'}{u} h_{ij}-\frac{1}{2}h'_{ij}\Big) ~,\cr
\Gamma^{\phi}_{\theta\phi} &=\cot\theta-\frac{u'}{u} ~,\quad 
\Gamma^{i}_{\theta j}=\frac{1}{2} h^{ik}h'_{jk} -\frac{u'}{u}\delta^{i}_{j}~,
\end{align}
where the prime denotes derivative with respect to $\theta$. The non-zero components of the Ricci tensor of $g_X$ are
\begin{align}\label{RicX}
R_{\theta\theta}&= -\frac{1}{2}{\rm Tr}(h^{-1}h'')+\frac{1}{4}{\rm Tr}(h^{-1}h'h^{-1}h')+\frac{1}{2}\frac{u'}{u}{\rm Tr}(h^{-1}h')+1+\cot\theta\frac{u'}{u}+(d+1)\bigg[\frac{u''}{u}-\bigg(\frac{u'}{u}\bigg)^2\bigg]~,\cr
R_{\theta i}&= \frac{1}{2}h^{jk}\nabla_j h'_{ik} -\frac{1}{2}h^{jk}\nabla_i h'_{jk}~,\cr
R_{\phi\phi}
&=\frac{\sin^2\theta}{n^2}\bigg[1+\frac{u''}{u}+\bigg(\frac{u'}{u}-\cot\theta\bigg)\bigg(\frac{1}{2}{\rm Tr}(h^{-1}h')-(d+1)\frac{u'}{u}\bigg)\bigg]~,\cr
R_{ij}
&=R_{ij}(h)-\frac{1}{2R_0^2}h''_{ij}  +\frac{1}{2R_0^2} (h'h^{-1}h')_{ij}- \frac{1}{2R_0^2}\bigg[\cot\theta-d\,\frac{u'}{u}+\frac{1}{2}{\rm Tr}(h^{-1}h')\bigg] h'_{ij}\cr 
&\hspace{4.5cm}+\frac{1}{R_0^2}\bigg[\frac{u''}{u}-(d+1)\bigg(\frac{u'}{u}\bigg)^2+\cot\theta\frac{u'}{u}+\frac{1}{2}\frac{u'}{u}{\rm Tr}(h^{-1}h')\bigg] h_{ij}~. 
\end{align}
In the above, we have introduced the simplified notation ${\rm Tr}(h^{-1}h'')=h^{ij}h''_{ij}$, ${\rm Tr}(h^{-1}h'h^{-1}h')=h^{ij}h'_{ik}h^{kl} h'_{jl}$, and $(h'h^{-1}h')_{ij}=h'_{ik}h^{kl}h'_{jl}$.
Also, in the last equation, we have explicitly indicated that $R_{ij}(h)$ is Ricci tensor of the metric $h$. 
%
{\footnotesize
\bibliographystyle{JHEP}
\bibliography{AdSRefs}
}
\end{document}